\documentclass[preprint, twocolappendix]{emulateapj}

\newcommand{\mynew}[1]{{#1}}
\usepackage{rotating}
\usepackage[inline]{enumitem}
\usepackage{color}
\usepackage{ifthen}
\usepackage{amsmath}
\RequirePackage{xspace}
\def\ifundefined#1{\expandafter\ifx\csname#1\endcsname\relax}
\ifundefined{ensuremath}\def\ensuremath#1{\relax\ifmmode{#1}}
\else${#1}$\fi\else\relax\fi

\newcommand{\unit}[1]{\ensuremath{\,\, \mathrm{#1}} \xspace}  
\newcommand{\mycm}{\unit{cm}}
\newcommand{\mys}{\unit{s}}
\newcommand{\myG}{\unit{G}}
\newcommand{\myK}{\unit{K}}
\newcommand{\mykm}{\unit{km}}
\newcommand{\mykms}{\unit{km/s}}
\newcommand{\mygccm}{\unit{g/cm^{3}}}
\newcommand{\myergg}{\unit{erg/g}}
\newcommand{\mypu}{\unit{dyn/cm^2}}
\newcommand{\myviscu}{\unit{cm^2/s}}
\newcommand{\mysinv}{\unit{s^{-1}}}

\newcommand{\ldquot}{``}
\newcommand{\rdquot}{"}

\newcommand{\myg}{\mathrm{g}}
\newcommand{\mygvec}{\mathbf{g}}

\newcommand{\myBvec}{\mathbf{B}}

\newcommand{\myihat}{\widehat{\mathbf{i}}}
\newcommand{\myjhat}{\widehat{\mathbf{j}}}
\newcommand{\mykhat}{\widehat{\mathbf{k}}}
\newcommand{\myyhat}{\widehat{\mathbf{y}}}
\newcommand{\myyzhat}{\widehat{\mathbf{yz}}}
\newcommand{\myzhat}{\widehat{\mathbf{z}}}
\newcommand{\myMCh}{\ensuremath{M_{\rm Ch}}}
\newcommand{\myDm}{\Delta m_{15}}

\newcommand{\myHe}{He}
\newcommand{\myC}{{^{12}{C}}}
\newcommand{\myO}{{^{16}{O}}}
\newcommand{\myCO}{{C}\text{-}{O}}
\newcommand{\myNi}{{^{56}{Ni}}}

\newcommand{\myfuel}{\rm{fuel}}
\newcommand{\myprod}{\rm{prod}}
\newcommand{\myproduct}{\rm{product}}

\newcommand{\mymathnuc}[1]{\mathcal{#1}}
\newcommand{\myA}[1]{\mymathnuc{A}_{#1}}
\newcommand{\myX}[1]{\mymathnuc{X}_{#1}}
\newcommand{\myY}[1]{\mymathnuc{Y}_{#1}}
\newcommand{\myQ}{\mymathnuc{Q}}

\newcommand{\myod}[2]{\frac{ d #1}{ d #2}}
\newcommand{\mypd}[2]{\frac{ \partial #1}{ \partial #2}}

\begin{document}

\title{Magneto-Hydrodynamical  Effects on Nuclear Deflagration Fronts in Type Ia Supernovae}
\author{Boyan {Hristov} \altaffilmark{1,2}}
\author{David ~C. {Collins} \altaffilmark{2}}
\author{Peter {Hoeflich} \altaffilmark{2}}
\author{Charles ~A. {Weatherford} \altaffilmark{1}}
\author{Tiara ~R. {Diamond}\altaffilmark{3}}

\altaffiltext{1}{Department of Physics, Florida A\&M University}
\altaffiltext{2}{Department of Physics, Florida State University}
\altaffiltext{3}{NASA Goddard Space Flight Center, Greenbelt, MD 20771, USA}

\begin{abstract}

This article presents the study of the effects of magnetic fields
on non-distributed nuclear burning fronts as a possible solution
to a fundamental problem for 
the thermonuclear explosion of a Chandrasekhar mass ($\myMCh$) white dwarf (WD), 
the currently favored scenario for the majority of Type Ia SNe (SNe~Ia).
All existing 3D hydrodynamical simulations predict
strong global mixing  of the burning products
due to Rayleigh-Taylor (RT) instabilities,
which is in contradiction with observations.
As a first step and to study the flame physics
we present a set of computational magneto-hydrodynamic (MHD) models
in rectangular flux tubes, resembling
a small inner region of a WD.
We consider initial magnetic fields up to $10^{12} \myG$
of various orientations.
We find an increasing suppression of RT instabilities starting at about $10^9 \myG$.
The front speed tends to decrease with increasing magnitude up to about $10^{11} \myG$. 
For even higher fields new small scale finger-like 
structures develop, which increase the burning speed
by a factor of 3 to 4 above the field-free RT-dominated regime.
We suggest that the new instability may provide 
sufficiently accelerated energy production 
during the distributed burning regime
to go over the Chapman-Jougey limit and trigger a detonation.
Finally we discuss the possible origins of high magnetic fields
during the final stage of the progenitor evolution or the explosion.

\end{abstract}

\maketitle

\newif\ifpdf
\pdftrue

\section{Introduction}
\label{intro}

Type Ia supernovae (SNe Ia) are spectacular explosions at the end of the life of stars.
Similar explosion energies, spectra, light curves (LC) and LC decline rates,
$\myDm$, \mynew{established them as} standard candles capable of measuring the Universe 
at the largest cosmological scales \citep{p93}.
\mynew{Attempts to calibrate a brightness-decline relation began
as early as \citet{pskovskii77}}.
Recently \citet{li03,li11}, \citet{foley13} and \citet{Howell2006} identified the Iax 
and the super-Chandrasekhar mass  subclasses, which further complicates our understanding of these objects.
\mynew{A robust predictive model of SNe Ia first principles is in order.}

It is widely accepted that a Type Ia supernova is a thermonuclear explosion
resulting from a binary system, of which one star is 
necessarily a degenerate carbon-oxygen ($\myCO$) white dwarf (WD)
\citep{hf60} \mynew{near the Chandrasekhar mass (\myMCh).}
Current research considers various progenitor configurations and 
final outcomes.
Depending on whether or not both stars are WDs, the progenitor system  
is called double degenerate (DD) or single degenerate (SD).
In addition \mynew{to progenitor configuration,} proposed scenarios are distinguishable by the ignition mechanism
and other characteristics.
In the case of a DD progenitor system, a dynamical merger or a violent collision
between the WDs is capable of releasing enough heat to trigger an ignition.
This process can end up as a SN Ia, a highly magnetized WD (MWD), 
or an accretion-induced collapse (AIC)
\citep{iben,webbink84,benz90,rasio94,segretain97,%
yoon2007,WMC09,WCMH09,loren09,isern11,pakmor10}.
Another class of SNe Ia scenarios is the double detonation  
of a sub-$\myMCh$ WD with accretion from a helium ($\myHe$) companion.
A detonation in the surface helium layer causes 
a secondary detonation in the core
\citep{wwt80,n82,livne1990,woosley94,hk96,Kromer2010,WK2011}.
Finally there is the $\myMCh$ explosions scenario, where the WD progenitor 
accretes material from a companion star and nuclear surface burning to C/O
leads to an increase of the WD mass.
With increasing  WD mass the electron gas in the central region becomes increasingly
relativistic, which leads to faster compressional heat release, a raise of the central 
temperatures, and the triggering of a central C/O deflagration front
when the mass of the progenitor approaches $\myMCh$.
\citep{hf60,Sugimoto1980,nomoto82,2002ApJstein,Piersanti2004}.
It is likely that the dynamical merger, $\myMCh$ explosion, and double-detonation channels all contribute to the SN~Ia population  
because of the \ldquot stellar amnesia\rdquot effect (\citet{hoeflich06}, and references therein).
This can happen in either a SD system, 
where the donor star is a main-sequence star, a red giant, etc., 
or in a DD system with another WD being the donor 
\citep{WI73,Piersanti2004}.

 In this paper, we will focus on the $\myMCh$ explosion channel because of its 
consistency with a wide range of observations and their statistical properties.
From observations we learn that the ejecta of a typical SNe Ia
is made of chemical layers \citep[e.g.][]{barbon90,kirshner93,h95,fisher98,HGFS99by02,marion03,stehl04,tanaka11}.
The overall density structure shows \mynew{only}
small deviations from spherical geometry based on the continuum 
polarization \citep{howell01,maund10a,patat12} and based on  close to spherical supernovae remnants \citep{rest05,badenes06,fesen07,rest08}.

Despite the success of $\myMCh$ explosions, there are serious problems related to mixing by  the 
Rayleigh-Taylor (RT) instability during deflagration burning, prior to the phase of strong
expansion. \mynew{This instabilty results in}
 strong large-scale  mixing of the ejecta
\citep{khokhlov95,n95,livne99,rein99,gamezo03,roepke06,plewa2007}.
Although a layered chemical structure is partially restored during the detonation phase 
\citep{gamezo04,roepke12}, the predicted imprint of deflagration mixing is in contradiction to observations. 
\mynew{Several} observations of local SNe~Ia strongly suggest \mynew{that} a process \mynew{is
necessary} to partially suppress the dominant 
role of RT instabilities during the deflagration:
\begin{enumerate*}[label=(\alph*)]
    \item Direct imaging of the SNR s-Andromeda shows a large 
`Ca-free' core, indicative of high-density burning and limited mixing
\citep{fesen07,fesen15}.
\item \mynew{Spectral fits to observations of post-maximum spectra in normal-bright and subluminous SNe~Ia
are significantly 
degraded by injection of radioactive material into the S/Si layer
\citep[Figs.~12 and 14 in][]{HGFS99by02}.}
\item Flat-topped or stubby line profiles $1-2 \unit{years}$ after the explosion
    indicate
stable isotopes remain near the center after the initial phase of burning 
\citep{h04,motohara06,maedanature10,maeda,penney14,stritzinger14b,tiara15}.
        This is contrary to the mixing that would occur if RT full developed.
\end{enumerate*}
Good agreement has been obtained
with spherical models, whereas the flame physics is inherently
multi-dimensional, which 
causes extensive mixing and strongly degrades individual fits and statistical properties.
 For a detailed discussion, see \citet{h06,2017ApJ...846...58H}. 

Had there been a mechanism to suppress the RT instabilities 
in the early stages of the explosion, \mynew{these discrepancies} would be resolved.
This requires a new piece of physics that has previously been left out by the 
current multi-dimensional models for the deflagration phase. High magnetic fields
have been suggested as a solution \citep{h04,remming14,hristov16}.  \mynew{Several
points support this suggestion.}
\begin{enumerate*}[label=(\alph*)]
    \item From both theory and simulation,
\mynew{tension in the magnetic field} suppresses
RT instabilities parallel to the field and also secondary instabilities 
 perpendicular to the field \citep{chandra61book,stone07b,stone07a}.
\item Magnetic surface fields in a wide range of strengths have been observed in WDs 
\citep{2003AJ....125..348L,2007ApJ...654..499K,2012ApJS..199...29G,%
2013MNRAS.429.2934K,2014AJ....147..129S,%
2015SSRv..191..111F,%
2016MNRAS.455.3413K}.
While these observations indicate small fields, stronger fields are possible
        both within the core of the WD and at shorter time scales before the
        explosion (see Section \ref{discussion}).
    \item  Positron transport effects on light curves and spectral line profiles are expected  
        at late times for explosions without magnetic fields
        \citep[e.g.][]{milne99,penney14}.  However, these were not seen
in SN2011fe and SN2017j, which were observed for some three years past maximum light \citep{2017MNRAS.472.2534K,2017arXiv170401431Y}.
\item
\citet{penney14} used late-time near infrared line profiles to estimate magnetic
fields.
Though the number of SNe~Ia with late-time IR spectra is small, recent observations 
strongly suggest that initial high magnetic fields with strengths, $B > 10^6 \myG$ are
common \citep{h04,penney14,tiara15}, even assuming fields on small-scales comparable to the pressure scale height of the WD.
\end{enumerate*}
 
The effects of magnetic fields on the nuclear burning in
SNe Ia and in a stellar context have been studied in previous works.  
\citet{2012A&A...538A.115P} investigate the effects of magnetic fields 
on non-resonant thermonuclear reactions 
in the interiors of $\myCO$ WDs.
\citet{Kutsuna12} study laminar flames in magnetic fields
at densities that are typical for a $\myMCh$  WD.
\cite{ji13} simulated 
a MWDs merger on an axisymmetric cylindrical (2D) domain,
followed by \citet{2015ApJ...806L...1Z} in full 3D,
who obtained final fields of $>10^{10} \myG$.
\citet{ghezzi01,Ghezzi04} employ 1D semi-analytic flame models
and find that the flame speed is affected by suppression of RT instabilities
and additionally study the asymmetries of a deflagration front 
in a WD with a large-scale dipole field.

Theories about the origin of WD magnetic fields and in support of amplification
of the fields during different stages of the star or the binary system evolution
are discussed in the final section.

The article is organized in the following way. 
We describe our method in Section \ref{method} starting with arguments
about our choices of magnetic field strength and topology,
the initial conditions and boundary conditions, and the regime of burning.
The base MHD code, Enzo, is outlined in Section \ref{code}
and the \mynew{newly added} nuclear reaction model in Section \ref{burning}.
Our results are presented in Section \ref{results}. 
In Section \ref{discussion}, we 
discuss the results, as well as future work and 
current dynamo theories.
The latter is not part of this study but is important to justify
the use of high magnetic fields in the central region of the WD.
All equations are written in the {\em cgs} system of units.
Other units are used in plots and throughout the text on occasion.


\begin{figure*}
\includegraphics[width=.95\textwidth]{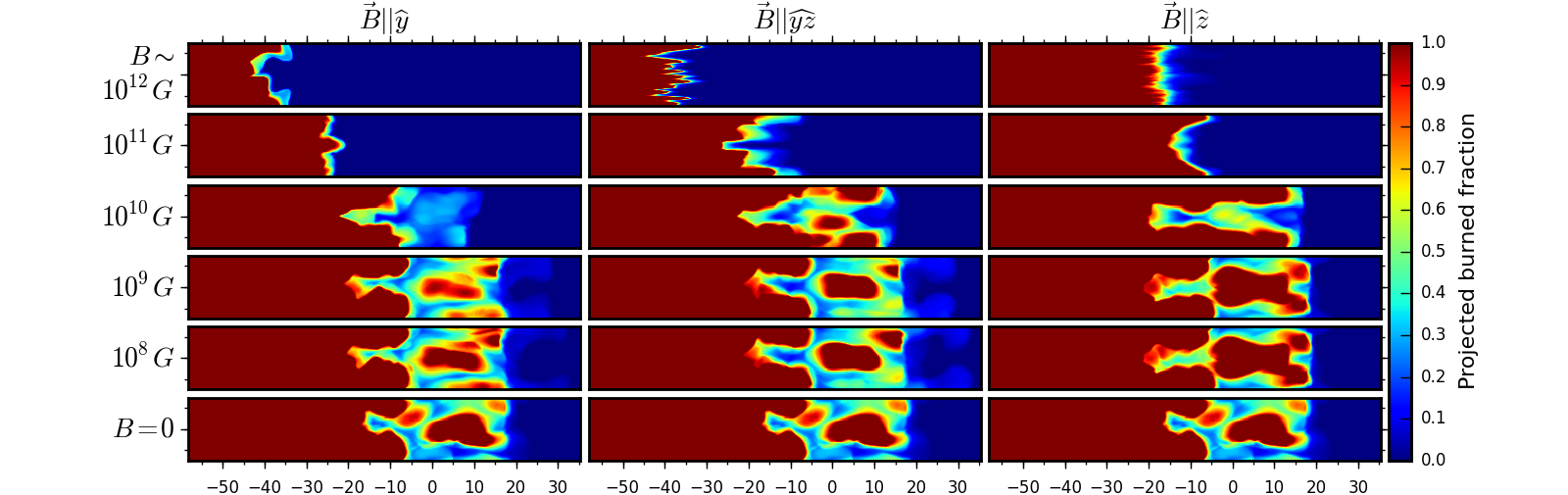}

(a) Fuel molar fraction at $t=0.6 \mys$ projected along the $y$-axis.

\includegraphics[width=.95\textwidth]{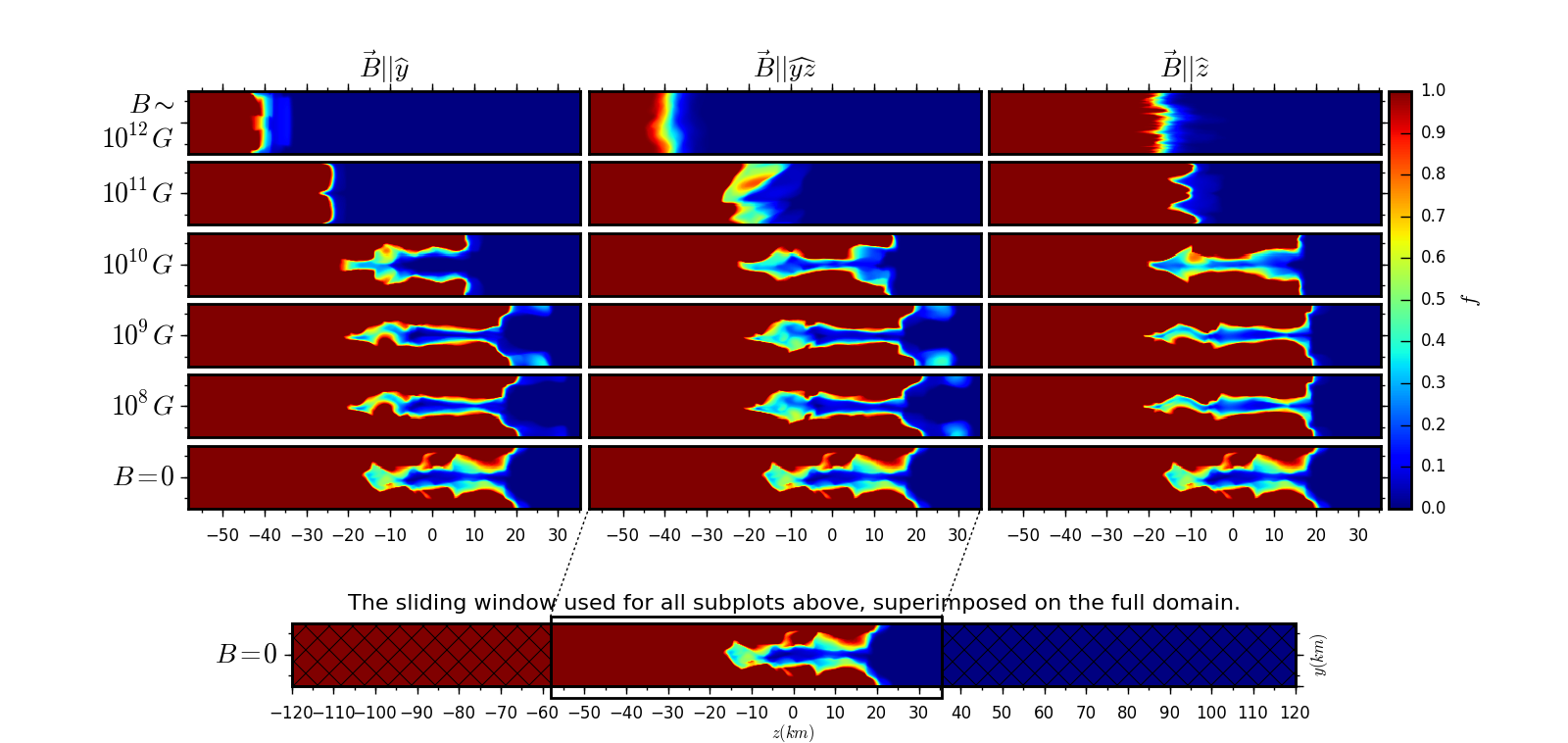}

(b) Fuel molar fraction at $t=0.6 \mys$ projected along the $x$-axis. 

\caption{
Effects of $B$ on the nuclear burning front.
At low or no field (bottom row and up) one  there is little difference in the behavior.
Turbulence dominates and cusps of fuel appear.
For models with $B_0=10^{10} \myG$ one can see some stabilization effects of the
front, most notably in Y10.
For models Y11 and Y12, where the field is in the plane of the front, the front becomes laminar.
For models YZ12 and Z12, the strong-field models with a component of the field in the direction of
the propoagation, we see another change in the burning properties.
Smaller modes along $z$ appear where the burning is accelerated.
The modes are stabilized by the strong magnetic field.
As a result, the fuel burns at \mynew{a higer} rate.
We suggest that this may be a mechanism to accelerate the front
over the Chapman-Jougey limit and trigger a transition to detonation.
See also Fig.~\ref{fig:qxqy}.
} 
\label{fig:fxfy}
\end{figure*}

\begin{figure*}
\includegraphics[width=.95\textwidth]{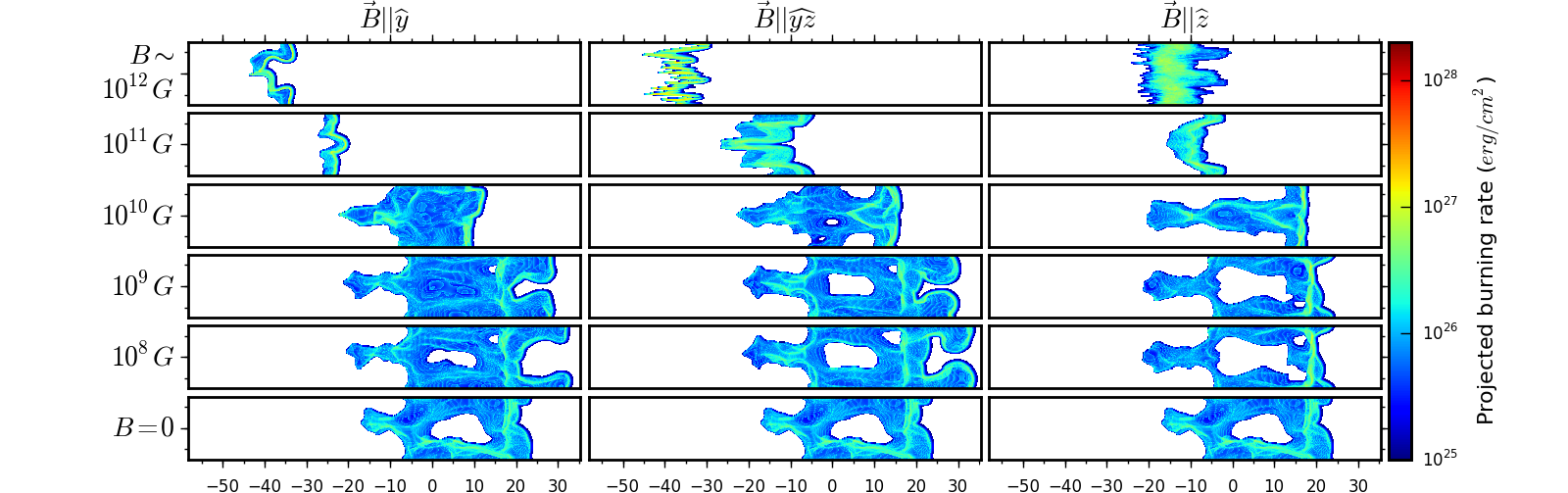}

(a) Energy production density rate projected along the $y$-axis.

\includegraphics[width=.95\textwidth]{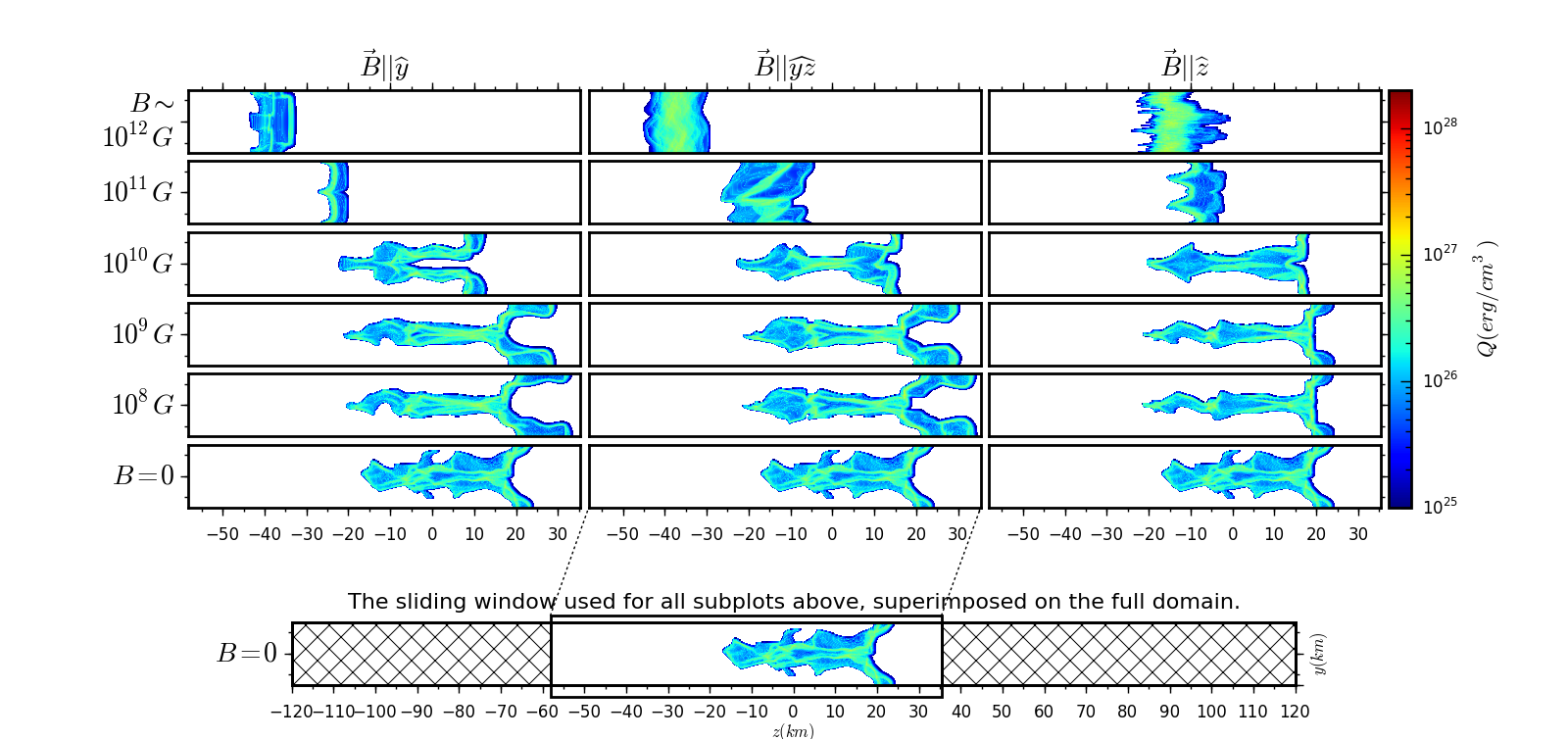}

(b) Energy production density rate projected along the $x$-axis.

\caption{Energy production density rate projections at $t = 0.6 \mys$. See also Fig.~\ref{fig:fxfy}.}
\label{fig:qxqy}
\end{figure*}

\section{Method}
\label{method}

To investigate the effects of magnetic fields on the nuclear burning
we ran 16 models identical in all
but the \mynew{magnitude and direction of the initial magnetic field.}  %
These runs are listed in Table \ref{tab:runmatrix}.
\begin{deluxetable}{lccc}
\tablecaption{Model (run) names, characterized by the size and orientation of the
initial $B$-field.}
\tablehead{
\colhead{\mynew{Initial} Magnitude} & 
\multicolumn{3}{c}{\mynew{Initial} Direction} \\\cline{2-4}\\
\colhead{(Gauss)} & 
\colhead{$\myBvec \parallel \myyhat$} & 
\colhead{$\myBvec \parallel \myyzhat$} & 
\colhead{$\myBvec \parallel \myzhat$}
}
\startdata
$ 1.4 \times 10^{12} $ & Y12 &  YZ12  & Z12  \\
$ 1.4 \times 10^{11} $ & Y11 &  YZ11  & Z11  \\
$ 1.4 \times 10^{10} $ & Y10 &  YZ10  & Z10  \\
$ 3.5 \times 10^{9}  $ & Y9  &  YZ9   & Z9   \\
$ 3.5 \times 10^{8}  $ & Y8  &  YZ8   & Z8   \\
$ 0                  $ & $B=0$ & ($B=0$) &($B=0$) \\
\enddata
\tablecomments{  The front propagates in the $\myzhat$ direction opposite to the gravitational acceleration, $\myg$. 
Figures \ref{fig:fxfy} and \ref{fig:qxqy} follow the same layout .}
\label{tab:runmatrix}
\end{deluxetable}
%
One model has no magnetic field and the rest 
\mynew{are initialized with a uniform magnetic field covering 5 orders of} magnitude
between $10^{8} \myG$ to $10^{12} \myG$,
and vary between 3 directions;
perpendicular to the WD radius, parallel to the WD radius and 
at $45^{\circ}$, labeled respectively 
$\myyhat$, $\myzhat$, and $\myyzhat$.
The main results can be sen in Figures \ref{fig:fxfy} and \ref{fig:qxqy}.
Figure \ref{fig:fxfy} shows the burned fraction for projections along $\hat{y}$
(panel a) and $\hat{x}$ (panel b).  Columns denote \mynew{initial} field orientation, and rows
denote \mynew{initial} field strength. Figure \ref{fig:qxqy} shows the burning rate for each
snapshot.  These figures are at $t=0.6 \mys$.  The bottom plot shows the position of
the burning front in the computational domain.  These will be
discussed further in Section \ref{results}.
To ensure the results are insensitive to the boundary, an additional simulation
was executed based on model  YZ10, but with twice the length of the domain.   This longer domain did not change the
results.
\mynew{
    We note that the \mynew{width} of the  domain imposes a maximum length of the possible modes,
which in turn allows for the existence of a sufficient magnetic field 
suppressing all linear perturbations \citep[see, for example, discussion in][]{stone07a}.
}

\subsection{White dwarf setup}
\label{wdsetup}

\begin{figure}
\includegraphics[width=0.45\textwidth]{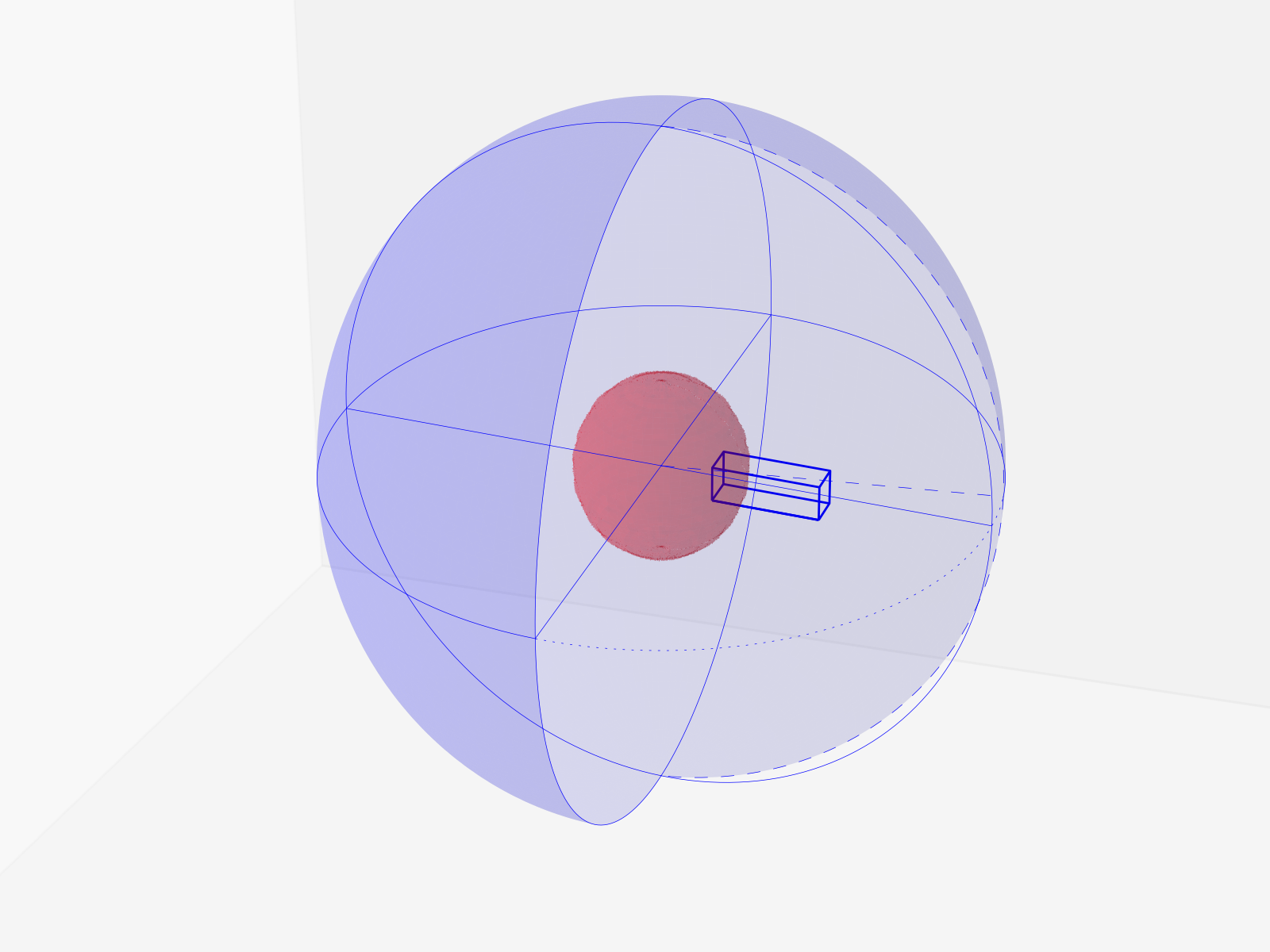}
\caption{
A sketch of the location of our flux tube within the white dwarf.  This is close
to the center, in the regime of non-distributed burning.  See Section
\ref{icbc} for more details.
}

\label{fig:sphere}
\end{figure}

Our work is in the context of a MWD undergoing a $\myMCh$-scenario explosion.
Model parameters were chosen to resemble the physical conditions of a WD after the onset
of the deflagration stage, long enough for RT instabilities to set up
and while the burning is still in the non-distributed regime. 

As \mynew{this is our first exploration of the impact of
magnetic fields on WD}, we use a simplified picture
in order to understand 
the physics of the burning front
in the presence of a magnetic field.
\mynew{We use} rectangular tubes with an initially uniform magnetic field,
embedded in the inner region of a WD
(Fig.~\ref{fig:sphere}.)
We solve the models numerically using
full 3D ideal MHD with non-distributed nuclear burning.
\subsubsection{Magnetic field magnitude}
\label{b-magnitude}

We consider $B$ magnitudes up to $10^{12} \myG$, 
which is close to those for a virialized WD
at $10^{13}-10^{14} \myG$
\citep{2015AN....336..851A,2017arXiv170200571F}. 
\mynew{As an order-of-magnitude estimate of the field strength, we can equate magnetic energy with
gravitational binding energy. For a density of $\rho = 10^8 \mygccm$ and a
radius of $1,600 \mykm$, using a binding energy of $G M^2/R$ and magnetic energy of
$B^2 R^3$, one finds
a field strength of $4\times 10^{12} \myG$.}

MWDs are commonly observed in the magnitude range $10^{3}-10^{9} \myG$ 
\citep{
2003AJ....125..348L,
2007ApJ...654..499K,
2012ApJS..199...29G,
2013MNRAS.429.2934K,
2014AJ....147..129S,
2016MNRAS.455.3413K%
}, where the observations suggest a real cutoff at the upper bound
and a sensitivity-limited lower bound
\citep{2015SSRv..191..111F}.
However, these observations are of non-exploding WDs, and
the observed magnetic field magnitudes are only on the surface of the WD.
Here, while we want to model a region close to the center and
after a central ignition has occurred, which is a regime not probed by these
observations.  Small-scale dynamo theory, as well as virial analysis, suggest that fields as large as
$10^{12}$G are possible in the conditions expected before the explosion.  This
will be discussed further in Section \ref{dynamo}.

\subsubsection{Field morphology considerations}
\label{b-geometry}

Field morphology can become very complex, as predicted by theory and observations
(see \citealp{2005ASPC..330..177R}, and discussion therein).
It is also uncertain whether the field 
is a dipole (or a multipole) on a large scale or turbulent,
or if there is a combination of fields at different scales.
We try to accommodate both cases by considering 
the order of the typical turbulent eddy, 
$l_{\rm turb} \sim 100 \mykm$, established in \citet{2002ApJstein}.
Furthermore, the domain needs to be long enough to 
allow the advancement of the burning front 
for about a second, as well as to
make the boundary effects remote enough.
With this in mind we made our computational domain 
$240 \times 15 \times 15 \mykm$ 
\mynew{($1024 \times 64 \times 64$ computational zones,
see also section \ref{nondb})}
.

\subsubsection{Initial and boundary conditions}
\label{icbc}

Immediately after a central ignition, RT instabilities would not develop
because close to the center of the WD the gravitational acceleration, $ \myg \simeq 0 $. 
We need to pick a later time when the burning front becomes RT-unstable.
In addition we are interested in the early evolution of the WD,
because plumes created early on have the most time to rise
and therefore would 
create the most mixing.
Such conditions are realized in a model in \citet{hoeflich06}
at time $t=0.1 \mys$ when the burning front has reached about $1,700 \mykm$ from the center of the WD.
At that time and location $\myg$ and $\rho$ change little 
within the span of the domain, $L_{z}$, 
so we initialize them with constants
$\myg = 1.9 \times 10^{9} \unit{cm/s^2}$ 
and $\rho_{0} = 10^{8} \mygccm$.
Additionally we keep $\mathbf{g}$ constant in time.

We place the burning front at $10 \mykm$ from the bottom 
and perturb its plane, so the initial front surface is described by
$
\zeta(x, y)|_{t=0}
=
A_{0} (
\cos ( k_{0}x )
)
+
\cos ( k_{0} y )$, where
$k_{0} = 2 \pi / \lambda_{0} 
$.
\mynew{For all simulations, $A_{0}=5\times 10^{4} \rm{cm}$ and $\lambda_{0}=4.8\times 10^5
\rm{cm}.$} We initialize the region below the burning front with completely burned 
material and the one above with unburned fuel.
Both regions have the same uniform initial density of $\rho_{0} = 10^{8} \mygccm$.
\mynew{
The initial specific internal energy is also uniform 
yet different on both sides of the burning front,
$3.3 \times 10^{18}$ and $0.4 \times 10^{18} \myergg$
in the burned and unburned regions respectively,
corresponding to initial pressures of $8.25 \times 10^{26}$ and $10^{26} \mypu$.
The initial velocity is zero everywhere.
}

\subsubsection{Non-distributed regime of burning}
\label{nondb}

At the burning front, with temperatures of $T \gtrsim 10^{9.5} \myK$
and densities of $\rho_{0} \approx 10^{8} \mygccm$,
the laminar front thickness
drops to the order of centimeters:
$l_{\rm flame} \sim 10^{-2}-10 \mycm$
\citep{timmes}.
This is much smaller than other length scales that an attempt to capture the details 
of the burning front would be a formidable computational task.
As a consequence, and 
with the burning being complete in this regime,
we cannot resolve any intermediate isotopes but only fuel and burning products.
We need a grid resolution, $\Delta x$, 
small enough to reproduce accurate MHD features,
but much coarser that the front scale length,
$\Delta x \gg l_{\rm flame}$, and we use one 
similar to the resolution in \citet{khokhlov95} (see \ref{b-geometry}).

We chose a laminar front speed of $10 \mykms$ also after \citet{khokhlov95},
which is slightly higher than the velocity in \citet{timmes}.
Additionally we used a quasi-1D model to measure 
the numerical laminar front thickness and speed.
This model is similar to the B0 model, except that 
there is no perturbation of the initial burning front surface,
as well as there is no gradient in initial pressure
($10^{26} \mypu$ everywhere).
The results from this simulation confirmed that the model laminar
velocity is $10 \mykms$. 
Furthermore the flame thickness in our simulation is 
stretched over 4-5 computational zones (about $1 \mykm$),
which is typical for shock-capturing schemes.

\subsection{Code}
\label{code}

For the data presented in this paper, we solve the ideal MHD equations
with nuclear burning using the adaptive mesh refinement (AMR) code Enzo
\citep{Enzo13} extended to MHD by \citet{Collins10}. The nuclear burning is a
new addition to the code, and described in Section \ref{burning}.  
The MHD solver uses the hyperbolic solver of \citet{Li08a}, the isothermal HLLD Riemann solver
developed by \citet{Mignone07},  and the Constrained Transport (CT) method of
\citet{Gardiner05}.  The simulations presented here were run with fixed
resolution.  Enzo has been used in a diverse array of astrophysical settings,
including star formation \citep{Abel02, Collins12}, supersonic turbulence
\citep{Kritsuk07}, x-ray gas in clusters, \citep{Bryan98}, and large-scale
structure \citep{Bryan99,OShea15}, and the epoc of reionization
\citep{Norman15}, among others.  \mynew{For simplicity, all simulations are run
with fixed resolution}

The MHD equations with burning are as follows:
\begin{equation} \label{eq:MHD1}
\mypd{\rho}{t}
+ \nabla \cdot
(\rho \mathbf{v})
= 0
\end{equation}

\begin{equation} \label{eq:MHD2}
\mypd{\rho \mathbf{v} }{ t}
+ \nabla \cdot
\left( 
	\rho \mathbf{v} \mathbf{v} 
	+ \mathbf{I} P 
	- \frac { \mathbf{B} \mathbf{B} }{ 8 \pi } 
\right)
=
- \rho \mathbf{g}
\end{equation}

\begin{equation} \label{eq:MHD3}
\mypd{E}{t}
+ \nabla \cdot
\left[
	(E + P) \mathbf{v} 
	-
	\frac 
	{ \mathbf{B} ( \mathbf{B} \cdot \mathbf{v} ) }
	{ 4 \pi }	
\right]
=
- \rho \mathbf{v} \cdot \mathbf{g} 
+ \dot{Q}
\end{equation}

\begin{equation} \label{eq:MHD4}
\mypd { \mathbf{B} }{ t}
+ \nabla \times ( \mathbf{v} \times \mathbf{B})
=0
\end{equation}

where $\mathbf{v}\mathbf{v}$ and $\mathbf{B}\mathbf{B}$ are 
the velocity and the magnetic field outer products and $\rho$, $\mygvec$, and $\dot{Q}$ are the density, 
the gravitational acceleration, 
and the energy production rate from the nuclear burning.
The total energy density  is
equal to the total of \mynew{the thermal, 
the kinetic and the magnetic energy densities},

\begin{equation} \label{eq:E-total}
E = e + \frac{ \rho v^2 }{ 2 } + \frac{B^2}{8\pi}
\end{equation}

and

\begin{equation} \label{eq:P-total}
P = p + \frac{B^2}{8\pi}
\end{equation}

is the total pressure, which is the thermal plus the magnetic pressure.

\begin{figure*}
\includegraphics[width=0.95\textwidth]{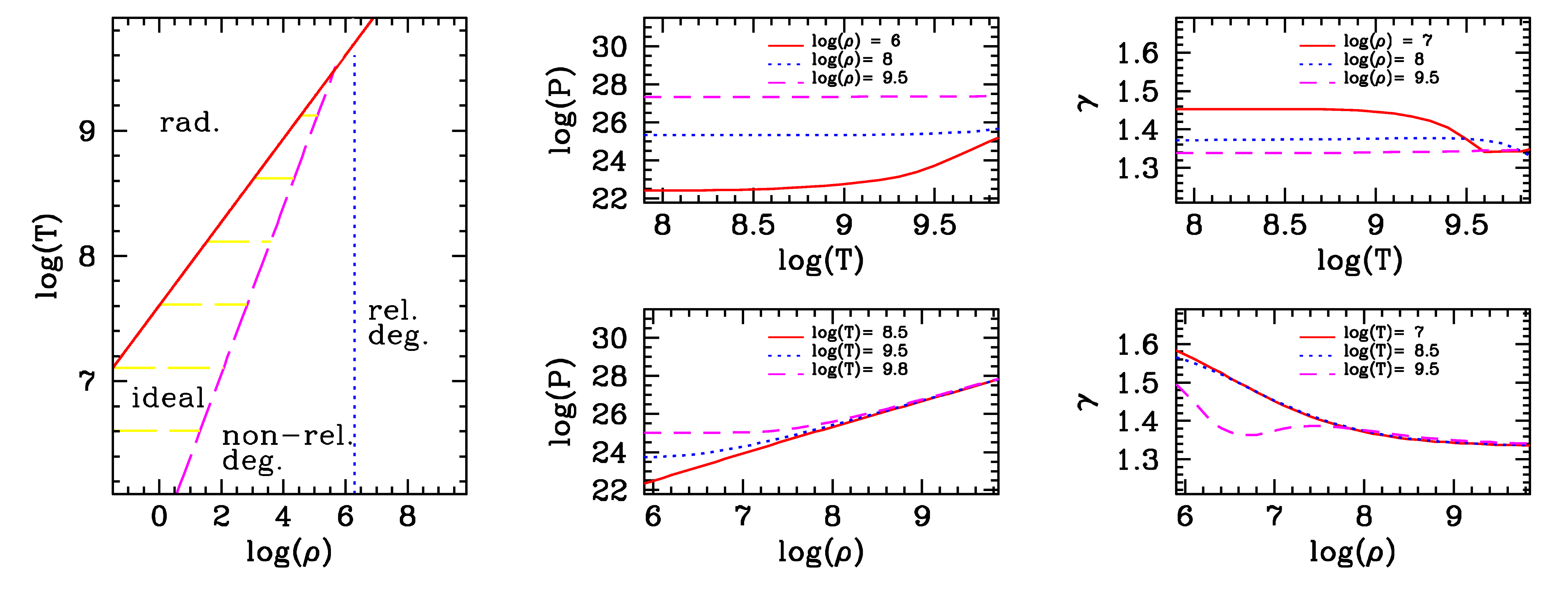}
\caption{
EOS of a WD \citep{hoeflich06}. 
The two right panels show that $\gamma$ depends relatively weakly
on $T$ and on $\rho$ in the non-distributed regime of burning
($T \ge 10^{8.5} \myK$ and $\rho \sim 10^{8} \mygccm $),
hence it can be taken to be a constant.
}
\label{fig:EOS}
\end{figure*}

A particular realization of a degenerate relativistic equation of state (EOS) 
from \citet{hoeflich06} is shown in Fig.~\ref{fig:EOS}.
Considering the right two panels, one can assume that 
at temperatures $T \gtrsim 10^{9.5} \myK$ 
and at densities in the range $\rho=10^{7}-10^{8} \mygccm$, the adiabatic index
$\gamma=\rm const$ is a reasonable approximation. 
Hence we close the system with the equations of state
for an ideal gas,
\begin{equation} \label{eq:EOS}
e = \frac{p}{\gamma - 1},
\end{equation}
where the adiabatic index $\gamma=1.4$ for these simulations.

\subsection{Burning}
\label{burning}

We model the nuclear burning and flame propagation after 
\citet{khokhlov95}, 
which takes advantage of the non-distributed regime
in a couple of ways.
Firstly it takes a grid resolution much coarser
than the flame scale length, as discussed in Section \ref{nondb}.
Then it approximates the real nuclear network with
one that has only two elements:

\begin{equation}
\label{eq:fueldproduct}
\myfuel \to \myproduct + \myQ
\end{equation}

where $\myQ$ is the specific energy produced from burning.
In our simulations the fuel consists of equal amounts by number
 of $\myC$ and $\myO$ with a mean atomic weight $\myA{\myfuel} \approx 14$,
and the product is $\myNi$.
\mynew{We} define the burned molar fraction, 
simply referred as burned fraction, $f$, as

\begin{equation} \label{eq:burned-fraction}
f = \frac { \myY{\myprod}  }{ \myY{\myfuel} + \myY{\myprod} }
\end{equation}

where $\myY{i}$ are the abundances.
Note that $ f \in {[0,1]} $, 
where $ f = 0 $ corresponds to pure fuel, 
and $ f = 1 $ means pure product. 
A nuclear deflagrating flame 
ignites unburned fuel
by heating through electron conduction. The model flame
spreads by diffusing the burned fraction, $f$:

\begin{equation} \label{eq:flame-diffusion}
\mypd{f}{t}
+
\mathbf{v} \cdot \nabla f
=
K \nabla^2 f
+
R(f).
\end{equation}
Here $K$ approximates the diffusion of burning material, thus energy to ignite
the up-stream fuel.  The instantaneous burning rate is then denoted by $R$.
The front becomes stretched over a few grid cells, 
as is typical for shock-capturing schemes.
The laminar speed velocity should be proportional to $\sqrt{KR}$.
We chose a value of $10\mykms$, which we confirmed in a 1D run.
According to \citet{remming16}, the laminar front velocity 
depends on the magnetic field strength and direction in addition to the fuel density.
However in our regime the difference the authors had calculated is
about 3\%, allowing us to assume constant values for $K$ and $R$.
Additionally we turn $R$ on and off locally depending on the burned fraction: 

\begin{equation} \label{eq:burning-rate}
R = R(f) =
  \begin{cases}
    R_{0}, \quad \text{ if } f_{0} \leq f \leq 1 ; \cr
    0, \quad \text{ if } 0 \leq f < f_{0} \\
  \end{cases}
\end{equation}
and $f_{0}=0.3$ as in \citet{khokhlov95}.

This delays the ignition of the fuel locally,
preventing the entire domain from been ignited 
in the first few time steps by non-physical numerical waves 
traveling faster than the physical laminar flame.
Finally, the burning energy is calculated as:

\begin{equation} 
\label{eq:Qdot}
\dot{Q} = \myQ \myod{\rho_{\myprod}}{t}
\end{equation}

\section{Results}
\label{results}

As described in Section \ref{method}, Figures \ref{fig:fxfy} and \ref{fig:qxqy} 
show the effect  of magnetic field on the propagation of nuclear burning in our
simulations.  There is a clear impact of the magnetic field on the speed and
morphology of the front in the strongest field simulations, but
one can see that 
all of the tested fields leave some imprint on the 
burning front
regardless of the initial magnitude or direction.
At low magnetic field, ($B \lesssim 10^{9} \myG$, and 
including no field at all, bottom row in those figures),
the \mynew{relative initial pressure (thus initial density) gradient is opposite
to gravity, 
making} the flow RT-unstable.
Large modes grow faster, as expected, and they would continue to rise.
The flame morphology becomes increasingly complex.
The burning front develops chaotic structure and 
is dominated by turbulence.
Small modes grow slower and are washed out by 
the flow shear or the diffusion 
by which the laminar front advances \citep[][see also eqn.
\ref{eq:flame-diffusion}]{khokhlov95}
The effective front width is $\sim 50-70 \mykm$.
Some pockets of fuel are formed.

Going up in magnitude to $10^{10} \myG$, 
we can recognize a trend of the effective width of the burning front
shrinking.
This is most pronounced in the case of a transverse field.
Turbulence is still prominent, and the front is comparable
to that of the lower $B$ cases.

At the highest magnetic field strengths, 
stabilization effects are well manifested.
The chaotic behavior is suppressed, and the front structure 
is closer to laminar.
The effective front width is now only on the order of $10 \mykm$.
For models Y11, YZ11 and Y12 the flame front closely follows the field lines.

\begin{figure}
\includegraphics[width=0.45\textwidth]{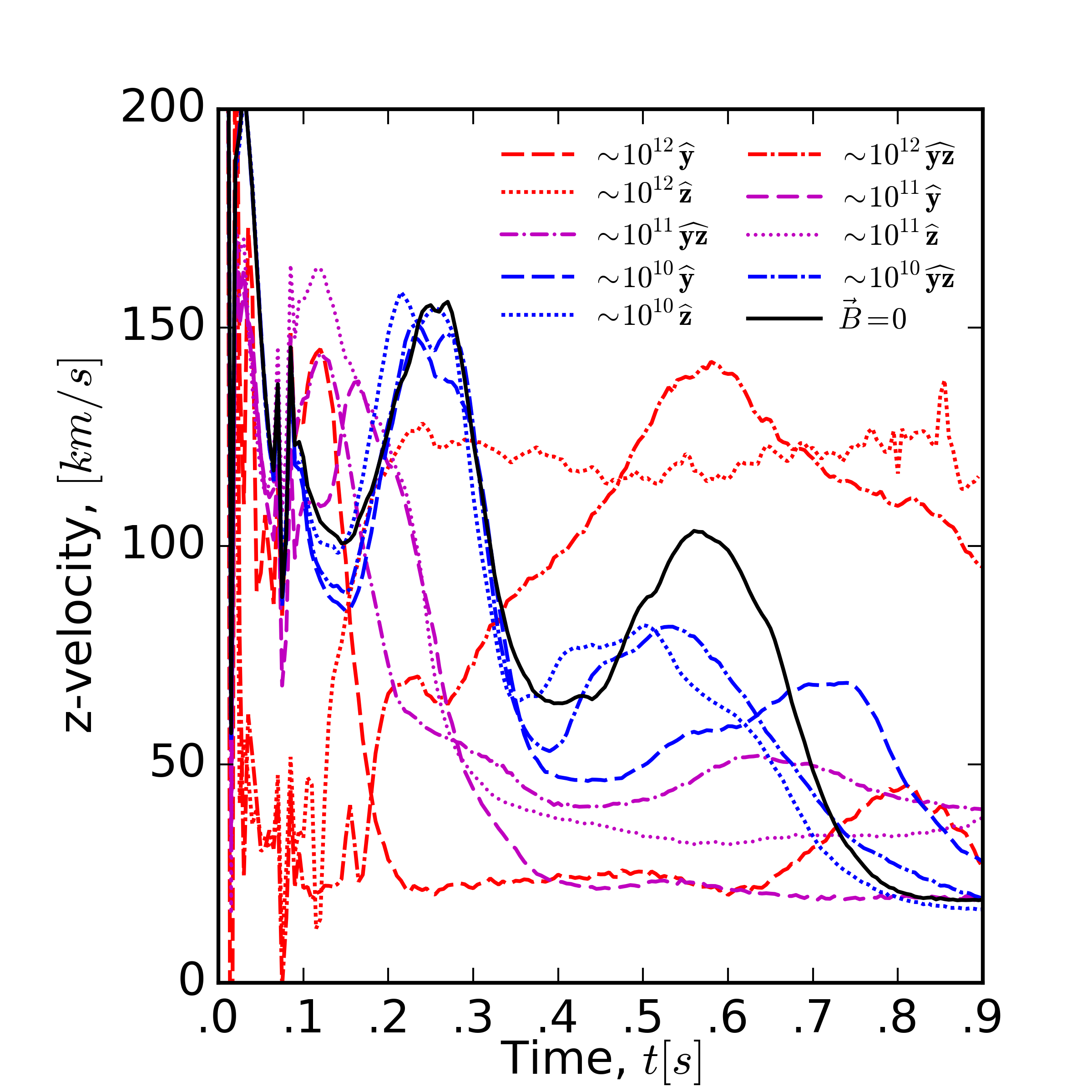}
\caption{
Flame speed in the frame comoving with the bulk flow.
The $10^{8} \myG$ and $10^{9} \myG$ profiles are omitted for clarity.
For most simulations, the velocity is decreased over the non-magnetized case by
the reduction of the surface area of the front.
The Z12 and YZ12 velocities
are larger than any other case as these regimes have new kind of instability as discussed in 
\mynew{ Section \ref{nonlinear}}.
}
\label{fig:vzfront} 
\end{figure}

\begin{figure}
\includegraphics[width=.45\textwidth]{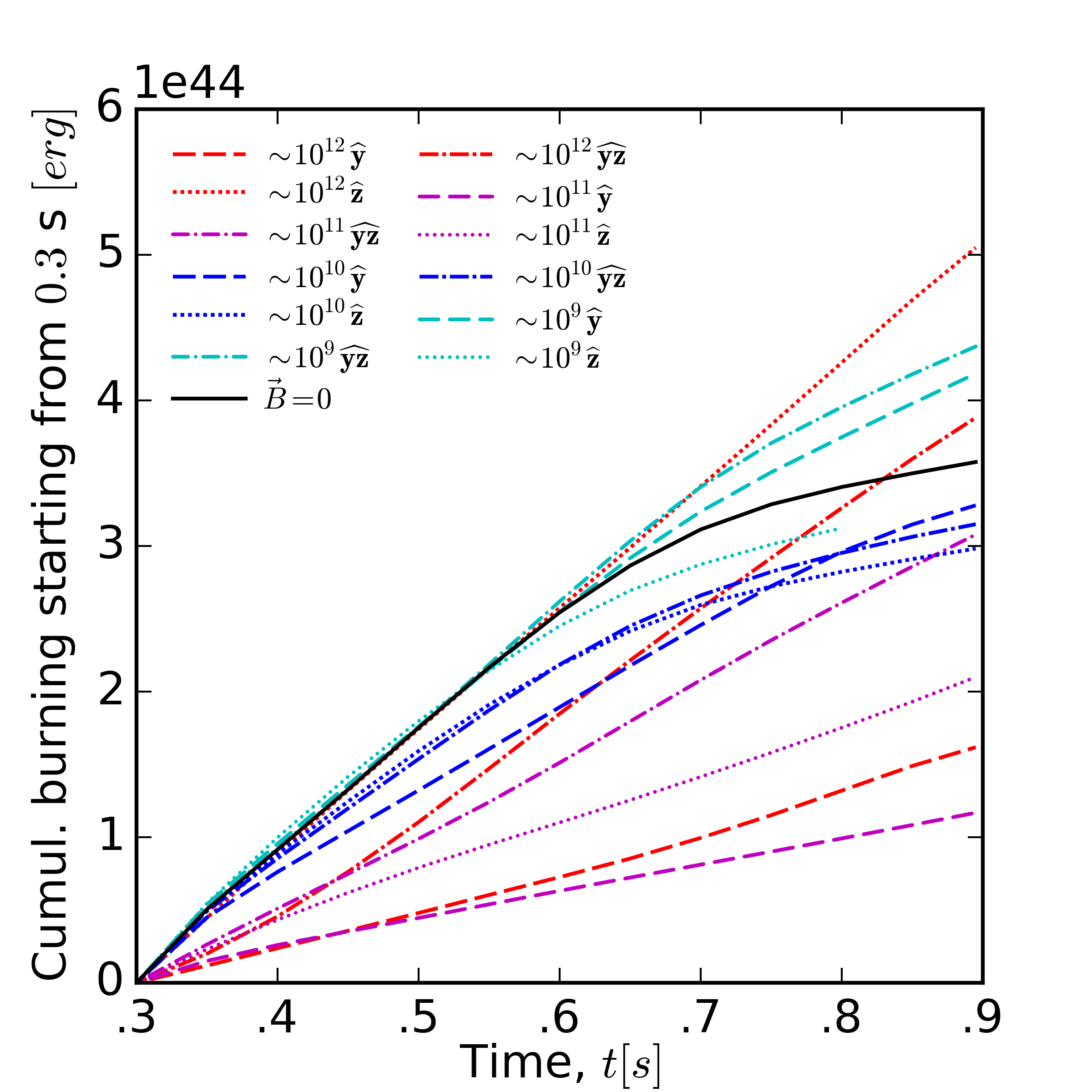}
\caption{
Cumulative burning energy, integrated from $t = 0.3 \mys$.
The $10^{8} \myG$ profile is not shown for clarity.
In order to remove the effects of the initial conditions,
seen in Fig.~\ref{fig:vzfront},
we integrate the burning rates starting right after the burst.
We note a trend for the rates to drop with increasing
field magnitude up to $10^{11} \myG$.
Beyond that models Z12 and YZ12
show the steepest slopes as well as steady state processes.
}
\label{fig:fz}
\end{figure}

For models Z12 and YZ12 we identify another change in the behavior.
The front becomes jagged with smaller peaks oriented along $z$.
In YZ12 these are seen only in the $y$-projection.
This behavior is mirrored in the front velocity and energy production, shown in Figures \ref{fig:vzfront} and \ref{fig:fz}, respectively.  In Figure
\ref{fig:vzfront}, the black line shows the velocity of the fiducial un-magnetized run.  After
an initial transient phase, as the instability sets in prior to t=0.3 s, it is
clearly seen that the moderately magnetized runs (blue lines at $10^{10}\myG$
and purple lines at $10^{11}$G) move slower due to the decreased surface area of
the front.  This decrease is also seen in the energy production rate, which can
be seen as a proxy for the surface area, see eqn. \ref{eq:burning-rate}.
In these runs, the instability is suppressed as expected by
perturbation analysis, discussed in Section \ref{perturb}.  For the 
largest field runs that also contain a component along the propagation
direction, however, the behavior is somewhat counterintuitive. These runs show a
significant \emph{increase} in the front velocity over all
other runs.  This is due to suppression of secondary Kelvin-Helmholtz
instability, see Section \ref{nonlinear}.  

\subsection{Linear Stability Analysis}
\label{perturb}

\begin{figure}
\includegraphics[width=0.45\textwidth]{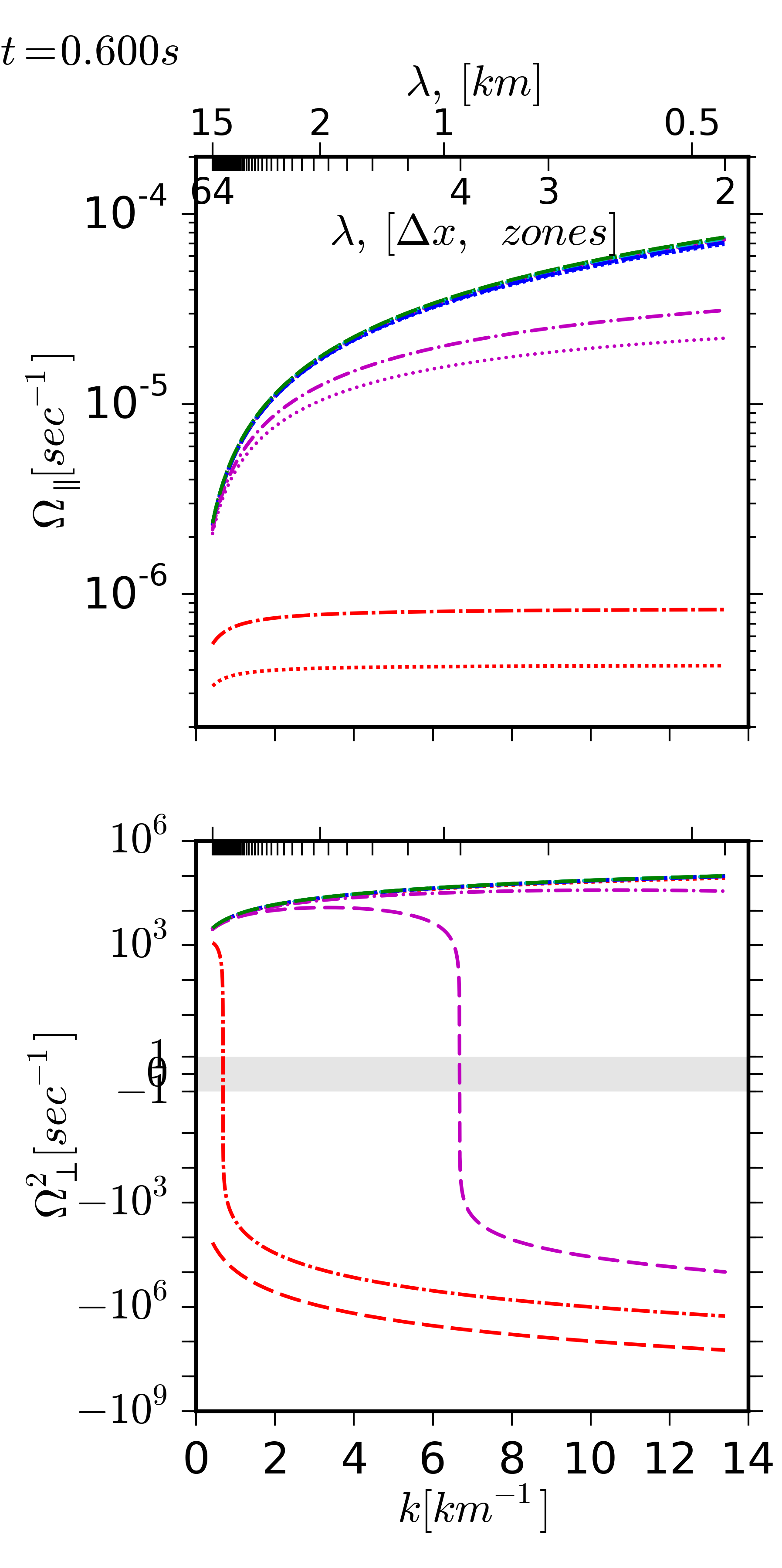}
\caption{
Growth rate of perturbation modes.
See legend on Fig.~\ref{fig:fz}. 
Negative values indicate stable modes.
The theoretical dispersion relations are used
for magnetic fields parallel (Eqs. \ref{eq:dispersion_par}--\ref{eq:dispersion_par3}),
and perpendicular (Eq.~\ref{eq:dispersion_perp})
to $\mygvec$ at $t=0.6 \mys$.
The wavelength range starts from 2 computational cells
up to the entire domain width.
For the perpendicular cases (bottom panel) we see that stable modes can be expected
only in three simulations, whereas 
all modes are stable in the Y12 run.
This is a rough agreement with the flows shown in Fig.~\ref{fig:fxfy},
especially in the $\myyhat$ models.
We also see stable surfaces along the magnetic field 
in the YZ11 run.
\mynew{
In the parallel cases (top panel) none of the simulations show stable modes.
and the smallest modes grow the fastest.
However the growth rate is suppressed so much that
 no growth will occur for the duration of the problem, 
$\sim 1 \mys$.}
}
\label{fig:disprel}
\end{figure}

The MHD equations (\ref{eq:MHD1}-\ref{eq:MHD4}) can be perturbed about an
interface subject to an acceleration, and the growth rates calculated, as is
done in standard Rayleigh-Taylor or Kelvin-Helmholtz instability studies.
The magnetic field can exert tension in the
front, and tend to  stabilize the flow.
Details are left to Appendix \ref{appxC}, but we will discuss the salient
results here.  The analysis assumes that the magnetic field is either parallel
or perpendicular to the gravity, $\mygvec$ and the position of the front is assumed to behave as 
$exp(i k_x x + i k_y y + \Omega t)$. Here, a negative  $\Omega^2$ indicates a stable mode.
The dispersion relation for the field perpendicular to $\mygvec$,
\mynew{presented on} the bottom panel of Fig.~\ref{fig:disprel}, shows that 
the only simulations with some stable modes in the linear regime are Y11, YZ11 and Y12.
All other orientations, at all modes, remain unstable.  This explains
the stark contrast between the top two rows and the other four in Figures \ref{fig:fxfy} and
\ref{fig:qxqy}.
\mynew{Having stable modes above a certain magnetic field strength
 is an effect of the finite domain size, as mentioned in the beginning of Section \ref{method}.}

\mynew{
When the field is parallel to the gravity, 
the top panel of Fig.~\ref{fig:disprel}
shows no stable linear modes regardless of the field strength.
This not a limitation of the domain size or the resolutiuon
but works for all wave numbers as proved in \citet{chandra61book}.
We also note that the growth rate is monotonic 
with respect to the size of the perturbation,
with the smallest modes growing fastest.
However any of the calculated growth rates
(considering all models, including $B=0$)
do not excede $10^{-4}$, 
which means that in this case
the duration of the problem, $\sim 1 \mys$
is too short for any linear perturbations to grow.
}

\subsection{Nonlinear Growth}
\label{nonlinear}

The linear stability is useful for guiding intuition, but the flow in question
is no longer in the linear regime.  

\begin{figure}
\includegraphics[width=0.45\textwidth]{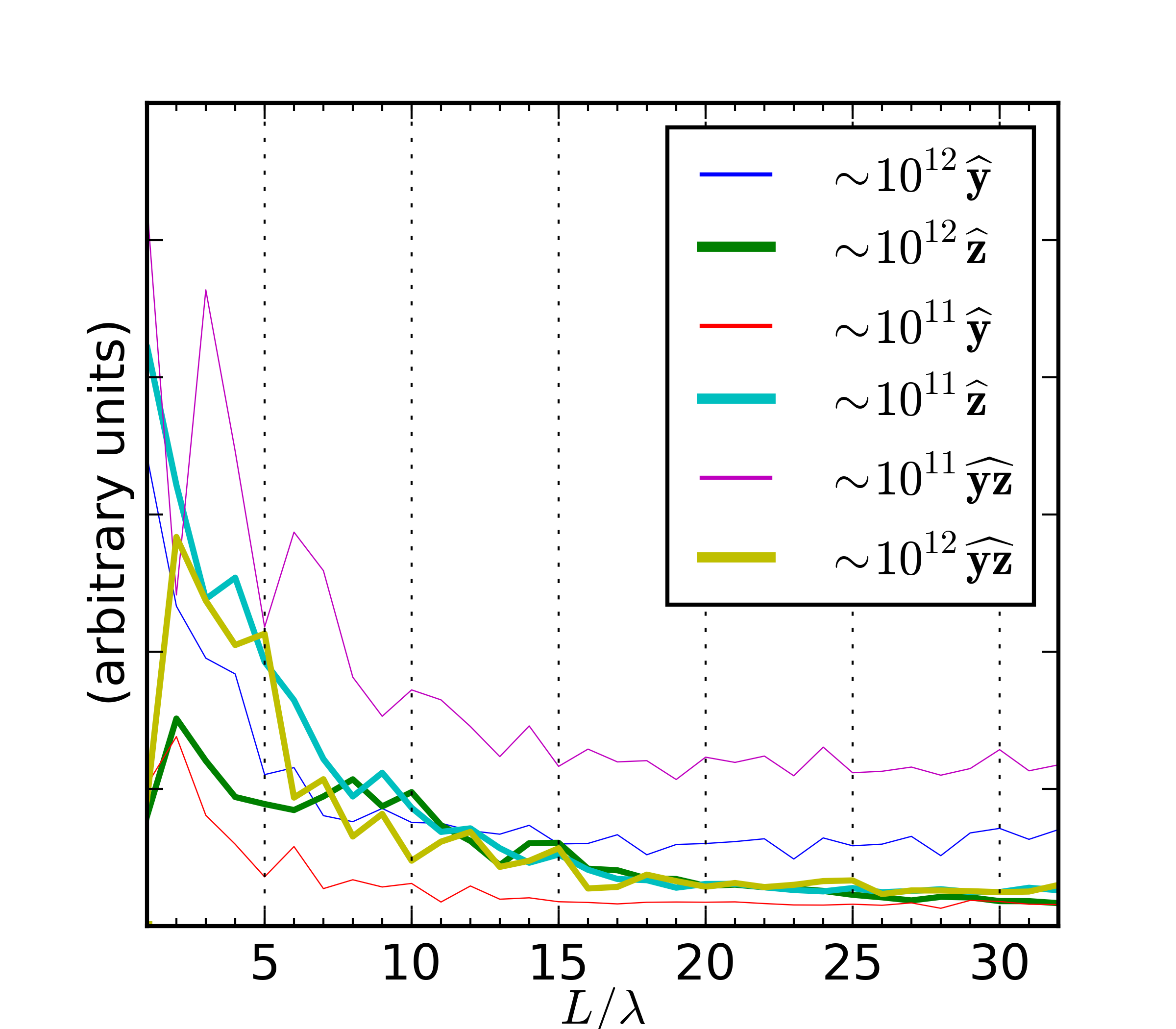}
\caption{
Fourier transformation of the flame as a function of $x$ and $y$ 
for the high-field simulations. 
All profiles are normalized so they
have maximum 1, in order to emphasize the distribution of power within
each simulation, rather than comparing powers between simulations.
The Z12 spectrum shows stronger features for
$L/\lambda = 8-15$ compared to the rest.
\mynew{
This is due to the
lateral magnetic support against secondary instabilities
once perturbations become elongated. }
Additionally all profiles become flat for $L/\lambda \ge 16$
with little power in that region, i.e. only modes larger than two
computational cells are prominent. 
This can be attributed to competition of advection vs. diffusion, but could be a resolution effect
as well \mynew{and needs to be studied further.}
}
\label{fig:fft}
\end{figure}

In order to quantify this impression
we applied a Fourier transform to 
the 95\% iso-surface of the burned fraction
in all $10^{11}$ and $10^{12}$ runs, shown in Fig.~\ref{fig:fft}.
This is a good method for the YZ12 and Z12 models, 
since the said iso-surface is a single-valued function,
$\zeta(x, y)$.
When the front has multiple points for the same $(x, y)$
we take the rightmost one.
Most profiles have the same power for the medium modes in the range 
$L/\lambda = 8 \-- 15$ zones.
However these are dominated by the larger modes 
 ($L/\lambda = 1 \-- 6$ zones)
and appear as noise.
The Z12 spectrum, on the other hand,
shows suppressed large modes to the same level \mynew{of} power as  
the medium modes.
Therefore we cannot consider the latter as noise.

\begin{figure}
\includegraphics[width=0.45\textwidth]{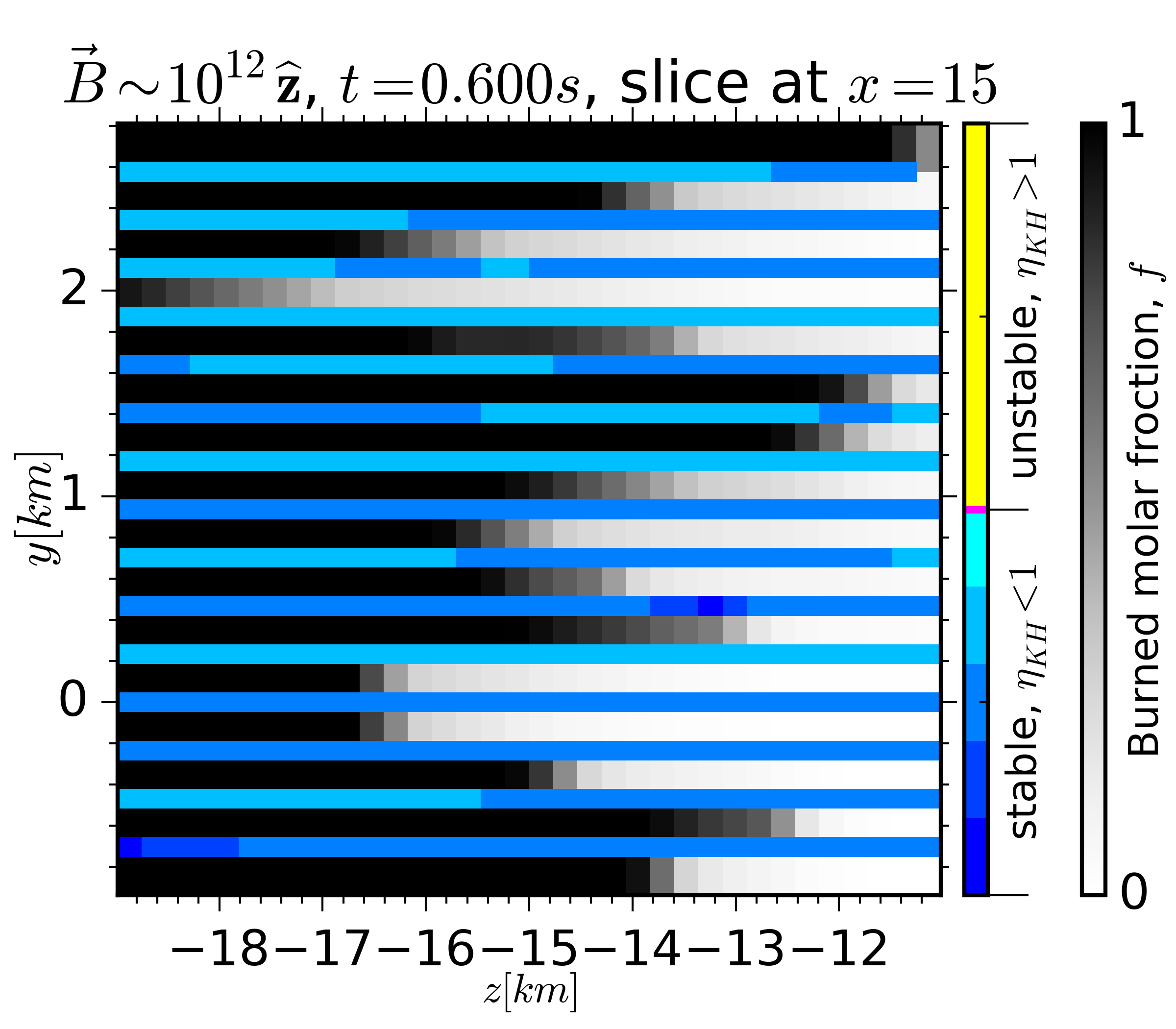}
\caption{
The ``Kelvin-Helmholtz number'' (Eq.~\ref{eq:khn}, yellow-to-blue scale) 
on the computational cells boundaries
superimposed on the burned fraction (gray scale)
for the Z12 model at $t = 0.6 \mys$.
Cold fuel sinks against the rising burned material,
creating shear on the vertical cell boundaries.
All lateral perturbations,
i.e. the secondary Kelvin-Helmholtz instabilities,
are suppressed by the magnetic field,
sustaining the finger formations.
}
\label{fig:khn}
\end{figure}

Furthermore the side walls of these instabilties
become almost parallel to $\myzhat$,
where the sinking fuel and the rising burned material
create shear at the boundary.
These conditions are right for developing 
Kelvin-Helmholtz instabilities, but
the magnetic field is now almost parallel to the
shear interface, adding surface tension 
and making the flow laterally stable.
The condition for Kelvin-Helmholtz stability, derived in the stability analysis
in the Appendix (in Eq.~\ref{eq:khn})
is visualized in a 1-zone slice in Fig.~\ref{fig:khn}. 
It holds for all zone boundaries parallel to $\myzhat$.
Somewhat similar effects to the RT fingers were observed 
by \citet{stone07b,stone07a} \mynew{ in lateral direction 
caused by magnetic fields perpendicular to the gravity.
Most importantly they consider non-recative fliuids
and observe bubble growth poportional to $t^2$.
In our models burning is expected to limit this growth in all regimes.
Specifically in Z12 and YZ12 models the height of the burning zone, 
once established, remains about the same through the simlulation.}

We also observe that all spectra become flat
for $L/\lambda > 15$ with the least power in those modes.
Since the lateral advection is suppressed we could attribute 
this effect to diffusion, 
however it could be something else, including numerical.

Fig.~\ref{fig:khn} shows fingers of fuel 
\mynew{with a width comparable to} the flame laminar thickness, 3--4 zones.
\citet{poludnenko11} and \citet{hicks15} found that when the radius of curvature at the 
bottom of the fingers becomes on 
the order of the laminar flame thickness or less,
the fuel burns at an accelerated rate.
The plots in Figures \ref{fig:vzfront} and \ref{fig:fz} confirm this.
We note that the the burning rates and the front speeds
tend to drop with magnitude and are smallest at about $10^{11} \myG$
due to the decrease of the front surface,
but in the Z12 and YZ12 models
the profiles become steeper again.

\section{Discussion}
\label{discussion}

We show that a magnetic field in the WD deflagration regime 
has an effect on nuclear burning for magnitudes $B \gtrsim 10^{10} \myG$.
At medium to high magnitudes 
we observe a front that is more organized compared 
to low magnetic field strength or no field at all.
The burning front becomes more laminar and less turbulent, and RT instabilities decrease.
The effective front thickness also decreases by a factor of a few.
One should ask whether medium fields are sufficient 
to bring significant change in the final outcome of the explosion,
given that a Type Ia deflagration is estimated  
to only last about 2 seconds.
The front speed and total burning rate 
tend to decrease as the field magnitude grows 
up to $10^{11} \myG$ regardless of the field direction.
\mynew{
Note that without steady state flows it is not meaningful 
to compare models one-to-one.
However it is clear that the YZ12 and Z12 models have the fastest fronts 
and the effects of the direction of $B$ are stronger.
}
When the longitudinal component of the field is strong,
the burning is amplified and the front speed 
is highest and larger than the pure hydrodynamical case by 
a factor of 3 to 4.

\subsection{Impact of Magnetic Fields}
These results suggest that strong magnetic fields might be 
missing from the picture of WD explosion physics
painted by current 3D models.
Suppression of large modes of RT instabilities 
in the early stages of the explosion
is necessary to prevent large scale mixing  
until the transition to detonation.
A second possible implication is related to the development of small scale
structures which result in an effective increase in the 
`surface' of the flame, which is the physical underpinning for
a significant increase of the burning rate compared to 
a pure hydrodynamical front.  
If burning rate is amplified by magnetic fields
it can lead to faster expansion and may freeze out 
 the RT plumes in the unburned environment,
consequently not allowing them rise to the surface,
which also means less mixing.

\mynew{
As discussed in the introduction, the delayed-detonation model is currently the favorite for $\myMCh$  explosions,
but DDT remains unexplained.
Possible processes based on the Zel'dovich mechanism are still under discussion.
In order to increase the nuclear burning rate and form a detonation front,
these require simultaneous mixing of burned and unburned matter on small scales 
over a significant volume. 
Possible physical causes for such mixing include mixing during a pulsational phase of the WD by
crossing shock waves produced in the highly turbulent medium, 
shear flows and instabilities in the regime of distributed burning which are discussed in \citet{poludnenko2016b}.
However, small scale fluctuations may be expected to prevent a DDT through mixing \citep{khokhlov97,n97}.
Another way to transition to detonation is by compression,
when the speed of the burning front reaches the so called Chapman-Jougey limit.
No solution for steady burning by deflagration exists 
above this limit, which 
for C/O rich mixtures is about 40\% of $v_s$  \citep{Bruenn}.
Based on 3D full-star simulations of deflagration fronts \citep[e.g.][]{gamezo03},
pure deflagration fronts reach about 10--15\% effective burning speed if full 
scale RT instabilities have developed over $\approx 2-3$ seconds. 
In \mynew{all simulations}, detonation in hydrodynamical models needs to be triggered ``by hand".
The RT stabilization effect of magnetic fields should additionally reduce the front surface,
thus the front speed leading to a decreased likelyhood of DDT by either mechanism.
Our results of small scale structure with an effective increase of the burning by factors of 3--5 in the YZ12 and Z12 models, 
}
may open an alternative mechanism for the DDT as a similar factor 
in the regime of distributed burning would bring the burning speed 
well above the Chapman-Jougey limit. Because the structure depends on the size and orientation of $B$, we must expect a wide 
variety and a dependence of the field morphology.  A final answer is beyond this paper and will be addressed in full-star simulations in future works.

\subsection{Magnetic Field Growth and Saturation}
\label{dynamo}
We have shown that $B$ fields larger than $10^{9...10} \myG$ will have a
significant effect on the nuclear burning fronts under WD conditions,
and may solve \mynew{some of the} current problems with $\myMCh$ explosions and  delayed-detonation models in particular.
As discussed in the introduction, some WD have magnetic fields but, for most WDs, the $B$ fields are 
\mynew{much smaller}
than those indicated by late-time observations of SNe~Ia.
Is it possible to produce high fields on relevant timescales prior to the explosion?
\mynew{Dynamos may be created}
during the accretion phase on time scales of $10^6 \unit{years}$, \mynew{or}
during the smoldering phase leading to the thermonuclear runaway on time scales of minutes \citep{2002ApJstein},
or during the hydrodynamical phase of instabilities during the explosion on time scales of seconds or less.
This leads us directly to $B$-field amplification in WDs and dynamo theories.

A seed magnetic field can be amplified by a dynamo operating in the convective
zone of a differentially rotating star, including WDs \citep{parker79,Thomas95,Brandenburg05}. 
In large scale dynamos, a toroidal field is produced by winding-up              
the poloidal component. The convective elements move upwards and
downwards, perpendicular to the
toroidal field lines, bending them and creating a new poloidal
component \citep{parker79}.
Large scale dynamos grow
with a typical timescale of the Alfv\' en times, $ t_A \approx R (4 \pi
\rho)^{1/2}/B \approx 300 \mys$ \citep{parker79}, which is \mynew{comparable to the final
final phase prior to the runaway. Therefore, this is not sufficient to increase the field
by several orders of magnitude.}

\mynew{
Alternatively, in the \emph{small scale} dynamo,  turbulence alone can amplify  
fields very quickly
\citep{Kazantsev68, Tayler73,Acheson78,Hawley96,Spruit02,Brandenburg05,Schekochihin07,braithwaite09,Beresnyak09,Duez10a,Duez10c}.  
Within the core of a white dwarf close to $M_{\rm{Ch}}$, turbulence will
undoubtedly play an enormous role. 
As the WD approaches $M_{Ch}$, the polytropic index, $\gamma \rightarrow
4/3$, and the stability against radial motions disappears.  Then kinetic energy
will exhibit a Kolmogorov cascade, with energy
distributed at all scales in a power law, $E=C k^{-5/3}$.  This power law
persists to the molecular viscous dissipation scale.  
Initially weak magnetic fields are stretched, twisted, and folded to increase
the strength of the field.
This mechanism will cause amplification
until the kinetic and magnetic energies at a given scale
are equal.  In the \emph{kinematic phase},  the magnetic energy is significantly
below the kinetic energy.
This occurs for large scales (small wave number) $k<k^*$, where the equipartition scale, $k^*$, is determined by the scale at which
amplification is balanced by dissipation and transfer to higher scales.  For
small scales (large wave number),
$k>k^*$, the magnetic field follows a Kolmorgorov cascade.   The equipartition
wavenumber, $k^*$, then decreases linearly with time \citep{Schekochihin07,
Beresnyak09}.
}

\mynew{
The details of this dynamo depend on the kinetic and magnetic dissipation
scales.  The kinematic viscosity and magnetic
diffusivity have been found to be $\nu=3.13\times
10^{-2} \myviscu$ and  $\eta=5.6\times
10^{-2} \myviscu$, respectively \citep{Nandkumar84,2017arXiv170201813I}.  
With a typical velocity and length of $V\simeq 16 \mykms$ and
$L~\simeq  200 \mykm$, respectively \citep[e.g.][]{Nonaka12},
 we find kinetic and magnetic Reynolds numbers
of $Re \equiv LV/\nu=10^{15}$ and $Rm\equiv LV/\eta = 5\times 10^{14}$,
respectively.  We also find viscous and magnetic length scales of 
$\ell_\nu=L Re^{-3/4}=10^{-4} \mycm$ and $\ell_\eta = 1.7\times
10^{-4} \mycm$.
Behavior of the turbulent dynamo depends on the magnetic Prandtl number, $Pm=\nu/\eta=0.5$ for this
system. 
In recent theoretical studies, \citep[e.g.][]{Schekochihin07, Beresnyak09,
Schober12}, it has been shown that the field can grow to energy equipartition at
a given scale in
an eddy turnover time at the smallest scale, essentially instantaneously, during
the kinematic phase.
\citet{Schober12} estimate this growth rate for both large and small $Pm$,
finding that the growth rate $\Gamma = 0.027 Rm^{1/2} V/L$, for low values of
$Pm$.   Conservative estimates for the $L,V$ and $Pm$ given
above \mynew{yield} a growth rate of
$5\times 10^{4} \mysinv$, or a doubling time of $2\times 10^{-5} \mys$.  Using
length and velocity scales  more appropriate for later phases of the WD
\citep[e.g.][]{2002ApJstein}, 
$V\simeq 200 \mykms$ and
$L\simeq  100 \rm{km}$, we find $\Gamma =  3 \times 10^6 \mysinv$. 
}

\mynew{
It has been
estimated \citep{Schekochihin07} that the timescale for the equipartition length
scale,
$1/k^*$, to reach the outer scale of the turbulence, $L$, is $t=L/U$.  
For our
first estimate, this is $12 \mys$, while for our second it is $0.5 \mys$.  
To estimate the final level of magnetic field when the dynamo ceases to function, \citet{Schober15} examine the feedback
mechanism of \citet{Subramanian99}. In this model, the magnetic diffusivity
increases with magnetic energy.  These authors find that in the small magnetic
diffusivity limit as much as 40\%\ of the total kinetic energy can be converted
to magnetic energy.  
This is similar to numerical findings,
 for example \citet{Haugen04b} find
30\% of the energy is magnetic, \citet{Cho00} who find 25\%, and
\citep{Beresnyak14} who finds 15\%.  
If  the peak magnetic energy is some
fraction, $f$, of the total, 
\begin{align}
    \frac{B^2}{8 \pi} = f \frac{1}{2} \rho V^2, \nonumber
\end{align}
and  $\rho=10^{9} \mygccm$ and $f=0.4$, we find a 
field of $10^{11} \myG$ and $1.4\times 10^{12} \myG$ for the two conditions
above.  
In a recent success of both laboratory plasma physics and theory,
\citet{Tzeferacos2018} used the Omega laser facility at the University of
Rochester to produce a $102 \unit{kG}$ magnetic field.  They found a value of the ratio of
magnetic energy to kinetic energy of $f=0.04$, which is somewhat lower than the
numerical results, but will still result in a substantial magnetic field in a
WD.
Ultimately the details of the evolution depend on the details of the simulation,
so full white dwarf simulations will be needed to specify the field for this
specific system.
}

\subsection{Future Work}
We want to emphasize that this study presents only a first step to address the MHD problem for reactive fluids.
Firm  conclusions for an exploding WD require many more questions to be addressed.
Here we will mention the limits of this study and questions which will be addressed in 
the near future by full-star models with our existing, more detailed nuclear networks. 
First, we treat the problem as if the WD was not expanding
during the simulation --
our gravitational acceleration is a constant
instead of decreasing with time.
Radial gradients in the gravitational acceleration,
the initial density, and the initial pressure were neglected.
Furthermore an evolved magnetic field would not be uniform,
so the flame will encounter varying magnitude and
direction as it advances. Moreover, the laminar diffusion speed 
will become directionally dependent. These effects will be studied in flux tubes using
our Monte-Carlo transport coupled within a Particle in a Cell scheme.
Finally, our nuclear network is too simple to carry out
simulations in a distributed regime of burning should one want to include the detonation phase as well.

High resolution simulations of a full star are required                                                                                
in order to overcome the limitations of the flux tube results
presented in this work.
Other scientific questions we want to address in the future are:
Are magnetic fields the missing physics in the current 3D models?
In particular, do magnetic fields make pre-expansion of the WD possible;
and if so, what is the mechanism -- is it 
by suppressing the RT instabilities,  by plume freeze-out on an accelerated background,
or in some other way?
What are observational signatures of the different field magnitudes 
and morphologies?
Can they lead to different outcomes of the explosions, i.e.
can different magnetic fields explain some of the diversity of SNe Ia?

\section*{Acknowledgements}
\addcontentsline{toc}{section}{Acknowledgement}

C.A. Weatherford and B. Hristov were partially supported by 
the Department of Energy, National Nuclear Security Administration, under award  {DE}{NA}0002630,
and the Nuclear Regulatory Commission, NRC-HQ-12-G-27-0091.
P. Hoeflich acknowledges support by the National Science Foundation (NSF) grant 1715133.
This work used the Extreme Science and Engineering Discovery Environment
(XSEDE), which is supported by the NSF grant 
ACI-1548562.  Simulations were performed on Stampede and Maverick at the Texas
Advanced Computing Center {\tt (https://www.tacc.utexas.edu)}, using XSEDE allocation
TG-AST140008.
Thank you to Braithwaite for helpful discussions on the dynamo theories and
the formation of ultra-high magnetic fields in WD,  
to S. Shore for helpful discussions on MHD instabilities.
Also thanks to Casey Mc Laughin and Prasad Maddumage, 
FSU Research Computing Center {\tt (https://rcc.fsu.edu)},
for help with technical problems.
Visualization was done using
\texttt{yt} \citep{yt}, 
\texttt{matplotlib} \citep{matplotlib}, and 
\texttt{numpy} \citep{numpy}.
Thanks to Nathan Goldbaum and Matthew Turk for help on \texttt{yt} issues.

\begin{appendix}
\addcontentsline{toc}{section}{Appendices}

\section{Mass and molar fraction definitions and identities}
\label{appxA}

Here we define some quantities used in the text
and give some conversions between them.
Consider the $\myfuel$ and $\myproduct$ mixture, 
which we'll label with 1 and 2, respectively, to ease the notation.
Obviously the partial mass densities 
add up to the total mass density:
\begin{equation}
\label{eq:rhoparts}
\rho = \rho_{1} + \rho_{2}
\end{equation}

\mynew{We then} define the mass, the molar, and the burned fractions,
$\myX{}$, $\myY{}$, and $f$:

\begin{equation}
\myX{i} \equiv \rho_{i} / \rho
,
\;\;\;\;\myY{i} \equiv \myX{i} / \myA{i}
\end{equation}

\begin{equation}
\label{eq:fdef}
f \equiv \frac{\myY{2}}{\myY{1} + \myY{2}}
\end{equation}

Clearly, the burned and the fuel fractions comprise the entirety of material,
i.e. the fuel fraction equals to $1-f$, should one need it.

Using the definitions above it is easy to derive the following identities.
The first two express the burned fraction, $f$, in terms of 
the total mass density, $\rho$,
and one of the partial mass densities, $\rho_{1}$ or $\rho_{2}$.
The following two equations show how to calculate the abundances,
$\myY{1,2}$, from the burned fraction, $f$.

\begin{equation}
\label{eq:fofrhorho2}
f 
=
\frac{ \myA{1} \left( \rho - \rho_{1} \right) }
{\myA{1}\rho - \left( \myA{1} - \myA{2} \right) \rho_{1}}
=
\frac{\myA{1}\rho_{2}}
{\myA{2}\rho + \left( \myA{1} - \myA{2} \right) \rho_{2}}
\end{equation}

\begin{equation}
\myY{1} = \frac{1-f}
{\myA{1} \left( 1 - f \right) + \myA{2} f}
,
\label{eq:Y2off}
\myY{2} = \frac{f}
{\myA{1} \left( 1 - f \right) + \myA{2} f}
\end{equation}

\section{Burning operator}
\label{appxB}
Provided that we know $\rho$, from Equations \ref{eq:rhoparts} and \ref{eq:fofrhorho2}, it
follows that we only need the mass fraction and one of the partial mass densities to determine the other one.
In our implementation of the new burning operator in Enzo
it is the product mass density, $\rho_{2}$, that is being stored and advected.
The burning operator comes last in the time cycle and comprises the following steps:
\begin{enumerate*}[label=(\roman*)]
\item find the burned fraction at the beginning of the cycle, using the second of Eq.~\ref{eq:fofrhorho2};
\item evolve the burned fraction using a 27-point stencil for the Laplacian;
\item update the density from Eq.~\ref{eq:Y2off}; and
\item update the energy from Eq.~\ref{eq:Qdot}.
\end{enumerate*}

\section{Linear Stability Theory}
\label{appxC}

We use analytic linear stability results from \citet{chandra61book}.
These come from linear stability analyses 
of the growth rate, $\Omega$, of
normal perturbation modes, 
$\mathbf{k} \equiv k_x \myihat + k_{y} \myjhat$
along the discontinuity interface,
so that a perturbation is proportional to
$\exp(ik_x x + ik_y y + \Omega t)$.
Modes with such dependance on time are stable when $\Omega^2 < 0$.

We split the magnetic field into two terms
, parallel and perpendicular to the gravity, i.e.
$\mathbf{B} \equiv B_z \mykhat + \mathbf{B}_{\perp}$,
where
$B_z \neq 0$ and 
$\mathbf{B}_{\perp} \equiv B_x \myihat + B_y \myjhat \neq \mathbf{0}$, respectively.
The dimensionless dispersion relation for $B_z$ is:

\begin{equation}
\begin{split}
\eta^3 
+
2 \kappa( \alpha^{1/2}_2 + \alpha^{1/2}_1 ) \eta^2
+
\kappa ( 2 \kappa + \alpha_1 - \alpha_2) \eta
- \\ -
2 \kappa^2 ( \alpha^{1/2}_2 - \alpha^{1/2}_1 )
= 0
\label{eq:dispersion_par}
\end{split}
\end{equation}

where

\begin{equation}
\eta \equiv \frac{\Omega_{\parallel}}{ \myg / V_{A,z} }
, \;
\kappa \equiv \frac{k_{\parallel}}{ \myg / V_{A,z}^2 },
\label{eq:dispersion_par2}
\end{equation}

\begin{equation}
V_{A,z}^2 \equiv \frac{ B_z^2 }{ 4 \pi (\rho_2 + \rho_1) }
, \;
\alpha_{1,2} \equiv \frac{ \rho_{1,2}}{ \rho_1 + \rho_2 },
\label{eq:dispersion_par3}
\end{equation}

and $\rho_1 < \rho_2$.
In this case no values of the parameters $B$ and $\rho_{1,2}$
yield stable modes.

We rewrite the dispersion relation for 
$\mathbf{B}_{\perp}$,
for the modes parallel to the magnetic field,
$\mathbf{k}_{\perp} \parallel \mathbf{B}_{\perp}$. 
Modes not parallel to $\mathbf{B}_{\perp}$ 
are ``less" stable since 
the negative term should be multiplied by
$\cos^2(\mathbf{B}_{\perp}, \mathbf{k}_{\perp})$.
\begin{equation}
\Omega^2_{\perp} = \myg k_{\perp} \left(
{
\frac{\rho_2 - \rho_1}{\rho_2 + \rho_1}
-
\frac{B_{\perp}^2 }
{ 2 \pi (\rho_2 + \rho_1) g }
k_{\perp}
}
\right )
\label{eq:dispersion_perp}
\end{equation}
Modes are stable when the right hand side of Eq.~\ref{eq:dispersion_perp}
is negative.

Finally we rewrite the condition for a mode to be Kelvin-Helmholtz stable as
\begin{equation}
\eta_{KH} 
\equiv 
\frac{\alpha_1 \alpha_2 \Delta v_z}{ 2 V_A,z^2 }
< 1.
\label{eq:khn}
\end{equation}

\end{appendix}


\bibliographystyle{apj}
\bibliography{article.bib,bh.bib}

\end{document}